\documentclass[a4paper,11pt]{article}
\usepackage{aaskaiid}
\usepackage{bm}

\newcommand{\hi}{\textrm{H\textsc{i}}}
\newcommand{\LCDM}{$\Lambda\mathrm{CDM}\;$}
\newcommand{\bk}{\bm{k}}

\usepackage{ctable} 

\newcommand{\wideims}{{\it Wide Band 1 Survey}}
\newcommand{\deeplowims}{{\it Deep SKA1-LOW Survey}}

\title{Cosmology with \hi\ Intensity Mapping}
\ShortTitle{\hi\ Intensity Mapping}

\author[1]{Laura Wolz}
\ShortName{L. Wolz et al.} 
\author[2,3]{Gabriele Autieri}
\author[4]{José Luis Bernal}
\author[5]{Maria Berti}
\author[1,9]{Philip Bull}
\author[6,7,8,9]{Stefano Camera}
\author[10]{Isabella P. Carucci}
\author[11]{Zhaoting Chen}
\author[12,1]{Steven Cunnington}
\author[13,14]{José Fonseca}
\author[1]{Keith Grainge}
\author[6,7]{Jiakang Han}
\author[9]{Wenkai Hu}
\author[15,23,24,9]{Dionysios Karagiannis}
\author[9,16]{Mario G. Santos}
\author[17, 9]{Marta Spinelli}
\author[18, 9]{Liantsoa F. Randrianjanahary}
\author[19]{Cora Uhlemann}
\author[2,20,21]{Matteo Viel}
\author[19]{Bernhard Vos-Ginés}
\author[22,9]{Jingying Wang}

\affiliation[1]{Jodrell Bank Centre for Astrophysics, Department of Physics \& Astronomy, The University of Manchester, Manchester M13 9PL, UK}
\emailAdd{laura.wolz@manchester.ac.uk}
\affiliation[2]{SISSA International School for Advanced Studies, Via Bonomea 265, 34136 Trieste, Italy}
\affiliation[3]{INFN – National Institute for Nuclear Physics, Via Valerio 2, I-34127 Trieste, Italy}
\affiliation[4]{Instituto de F\'isica de Cantabria (IFCA), CSIC-Univ. de Cantabria, Avda. de los Castros s/n, E-39005 Santander, Spain}
\affiliation[5]{D\'epartement de Physique Th\'eorique and Center for Astroparticle Physics,Universit\'e de Gen\'eve, Quai E. Ansermet 24, CH-1211 Genève 4, Switzerland}
\affiliation[6]{Dipartimento di Fisica, Universit\`a degli Studi di Torino, Via P.\ Giuria 1, 10125 Torino, Italy}
\affiliation[7]{INFN -- Istituto Nazionale di Fisica Nucleare, Sezione di Torino, Via P.\ Giuria 1, 10125 Torino, Italy}
\affiliation[8]{INAF -- Istituto Nazionale di Astrofisica, Osservatorio Astrofisico di Torino, Strada Osservatorio 20, 10025 Pino Torinese, Italy}
\affiliation[9]{Department of Physics \& Astronomy, University of the Western Cape, Cape Town 7535, South Africa}
\affiliation[11]{Institute for Astronomy, The University of Edinburgh, Royal Observatory, Edinburgh EH9 3HJ, UK}
\affiliation[12]{Institute of Cosmology \& Gravitation, University of Portsmouth, Dennis Sciama Building, Portsmouth, PO1 3FX, UK}
\affiliation[13]{Instituto de Astrof\'isica e Ci\^encias do Espa\c{c}o, Universidade do Porto, CAUP, Rua das Estrelas, PT4150-762 Porto, Portugal}
\affiliation[14]{Departamento de F\'isica e Astronomia, Faculdade de Ci\^{e}ncias, Universidade do Porto, Rua do Campo Alegre 687, 4169-007 Porto, Portugal}
\affiliation[15]{Dipartimento di Fisica e Scienze della Terra, Universit\`a degli Studi di Ferrara, Via G.\ Saragat 1, 44122 Ferrara, Italy }
\affiliation[16]{South African Radio Astronomy Observatory (SARAO), Cape Town, 7700, South Africa}
\affiliation[17]{Observatoire de la Cˆote d'Azur, Laboratoire Lagrange, Bd de l'Observatoire, CS 34229, 06304 Nice cedex 4, France}
\affiliation[18]{Astrophysics Research Centre \& School of Mathematics, Statistics and Computer Science, University of KwaZulu-Natal, Durban, 4041, South Africa}
\affiliation[19]{Fakult\"at f\"ur Physik, Universit\"at Bielefeld, Postfach 100131, 33501 Bielefeld, Germany}
\affiliation[20]{IFPU – Institute for Fundamental Physics of the Universe, via Beirut 2, I-34151 Trieste, Italy }
\affiliation[10]{ INAF, Osservatorio Astronomico di Trieste, Via Tiepolo 11, I-34143 Trieste, Italy}
\affiliation[22]{Shanghai Astronomical Observatory, Chinese Academy of Sciences, 80 Nandan Road, Shanghai 200030, China}
\affiliation[23]{INFN-- Istituto Nazionale di Fisica Nucleare, Sezione di Ferrara, via Giuseppe Saragat 1, 44122 Ferrara, Italy}
\affiliation[24]{Van Swinderen Institute, University of Groningen, Nijenborgh 3, 9747 AG Groningen, The Netherlands}

\abstract{The redshifted spectral emission from neutral hydrogen (\hi) at rest wavelength $\lambda=21\ \rm cm$ can be used as a tracer of large-scale structure and its evolution. Within the \hi\ intensity mapping method, sufficient signal-to-noise is achieved by integrating the line emission within large voxels over a wide sky area and line of sight depth which allows access to the largest scales of the matter distribution. The resulting tomographic maps usually feature low angular and high redshift resolution. The SKAO will be able to conduct \hi\ intensity mapping experiments observing up to $20,000\ \rm deg^2$ over a wide range of redshifts. For SKA-Mid, we will employ the array in a fast-scanning single-dish mode using Band 1 and 2 to access $0<z<3$, mapping an enormous volume with fast survey speed, allowing for the possibility of a commensal survey producing high angular resolution maps via the on-the-fly imaging of the visibilities. For SKA-Low, we will focus on deep observations to detect the \hi\ signal in a frequency band matching $3<z<6$. In this chapter, we will give an overview of \hi\ intensity mapping with the SKAO, including an outline of planned surveys, a discussion of observational challenges, and methodology for power spectrum methodology and forecasts. We present predictions on the constraining power on $\Lambda \rm CDM$ cosmology from \hi\ intensity mapping data via power spectrum, and other observables such as bi-spectrum and \hi\ stacking. We also demonstrate the synergy power of \hi\ intensity mapping with other cosmological surveys.}

\begin{document}
\newcommand{\actaa}{Acta Astron.} 
\newcommand{\araa}{ARA\&A} 
\newcommand{\aar}{A\&ARv} 
\newcommand{\aapr}{A\&ARv} 
\newcommand{\ab}{Astrobiol.} 
\newcommand{\aj}{AJ} 
\newcommand{\apj}{ApJ} 
\newcommand{\apjl}{ApJL} 
\newcommand{\apjs}{ApJSS} 
\newcommand{\ao}{Appl. Opt.} 
\newcommand{\apss}{Astro. \& Space Sci.} 
\newcommand{\aap}{A\&A} 
\newcommand{\aaps}{A\&AS.} 
\newcommand{\baas}{Bull. Am. Astron. Soc.} 
\newcommand{\caa}{Chinese A\&A} 
\newcommand{\cjaa}{Chinese J. A\&A} 
\newcommand{\cqg}{Class. Quantum Gravity} 
\newcommand{\gal}{Galaxies} 
\newcommand{\gca}{Geo. Cosmo. Acta} 
\newcommand{\icarus}{Icarus} 
\newcommand{\jcap}{JCAP} 
\newcommand{\jgr}{J. Geophys. Res.} 
\newcommand{\jgrp}{J. Geophys. Res. Planets} 
\newcommand{\jqsrt}{J. Quant. Spectrosc. Radiat. Transf.} 
\newcommand{\memsai}{Mem. SAIt} 
\newcommand{\mnras}{MNRAS} 
\newcommand{\nat}{Nature} 
\newcommand{\nastro}{Nat. Astron.} 
\newcommand{\ncomms}{Nat. Commun.} 
\newcommand{\nphys}{Nat. Phys.} 
\newcommand{\na}{New Astron.} 
\newcommand{\nar}{New Astron. Rev.} 
\newcommand{\physrep}{Phys. Rep.} 
\newcommand{\pra}{Phys. Rev. A} 
\newcommand{\prb}{Phys. Rev. B} 
\newcommand{\prc}{Phys. Rev. C} 
\newcommand{\prd}{Phys. Rev. D} 
\newcommand{\pre}{Phys. Rev. E} 
\newcommand{\prx}{Phys. Rev. X} 
\newcommand{\prl}{Phys. Rev. Let.} 
\newcommand{\psj}{Planet. Sci. J.} 
\newcommand{\planss}{Planet. Space Sci.} 
\newcommand{\pnas}{Proc. Natl Acad. Sci. USA} 
\newcommand{\procspie}{Proc. SPIE} 
\newcommand{\pasa}{PASA} 
\newcommand{\pasj}{PASJ} 
\newcommand{\pasp}{PASP} 
\newcommand{\rmxaa}{RMXAA} 
\newcommand{\sci}{Science} 
\newcommand{\sciadv}{Sci. Adv.} 
\newcommand{\solphys}{Sol. Phys.} 
\newcommand{\sovast}{Soviet Ast.} 
\newcommand{\ssr}{Space Sci. Rev.} 
\newcommand{\uni}{Universe} 

\maketitle
\section{Introduction}
Large-scale structure (LSS) surveys provide measurements of the evolution of the distribution of matter across cosmic time, allowing us to address fundamental and fascinating questions such as the nature of dark matter, dark energy, and gravitational physics on the largest length scales. 
In the recent decade, the field has moved into an era of precision cosmology deriving tighter constraints on cosmological model parameters. 
New results have been driven by technological advances in optical astronomy resulting in greater sensitivities, angular resolution precision, and the ability to survey large volumes over wide redshifts, most notably DESI \citep{DESI:2016fyo, DESI:2025fxa}, the Euclid satellite mission \citep{2020A&A...642A.191E}, and the upcoming Rubin Telescope of the LSST \citep{LSSTDarkEnergyScience:2018jkl, 2019ApJ...873..111I}.
However, advanced measurements also opened more questions as inconsistencies and discrepancies between different cosmological probes and surveys have been found, often termed \textit{tensions}, see \cite{2021CQGra..38o3001D} for a review. 
The understanding and untangling of observational systematics and astrophysical nuisance parameters is now more important than ever as any impactful new conclusions such as constraints on Dark Energy will be highly scrutinised and require confirmation through multiple surveys, probes and tracers.

The ubiquitous neutral hydrogen (\hi) gas has been suggested as an excellent tracer for galaxies and dark matter acting as a new and complementary probe of the LSS to optical observations. As \hi\ is the most fundamental element to baryonic processes, it has the additional benefit of tracing dark matter evolution and structure formation throughout cosmic history from dark ages to epoch of reionization and late-time astrophysics and cosmology. Radio telescopes can observe cold \hi\ gas via the redshifted 21 cm emission line, however, the line is faint in emission restricting resolved observations to  individual galaxies to the local Universe. The SKAO has the ability to continuously observe redshifted \hi\ emission with frequencies $0.05<\nu<1.42$ GHz resulting in redshift $27<z<0$ coverage albeit instrumental abilities will highly fluctuate for different frequencies and targeted \hi\ epoch. Cosmological LSS surveys using \hi\ as a resolved tracer for galaxies will only be probable for $z<0.4$ for AA4, the full SKA baseline design \citep{SKA:2018ckk}. 

The intensity mapping (IM) technique circumvents  limitation by averaging the collective emission from many unresolved galaxies within large ``voxels'', producing \hi\ intensity maps with lower angular resolution but retaining high spectral resolution \citep{Bharadwaj:2000av, Battye:2004re, Wyithe:2007rq, Chang:2007xk}. This observing strategy is akin to Epoch of Reionization (EoR) surveys, however the targeted \hi\ is now in shielded areas within dark matter haloes and hence acting as dark matter tracer rather than the inter-galactic gas. In this chapter, we consider \hi\ IM experiments where \hi\ traces LSS and hence targets constraining the cosmological model, a boundary usually loosely defined as $z=6$, though the exact end of the EoR is among the science goals of the SKAO. 

The \hi\ IM approach dramatically increases the survey speed compared to traditional galaxy surveys and achievable sensitivity on the large cosmological scales of interest to measure the Baryon Acoustic Scale which allows constraints on Dark Energy parameters across Cosmic times \citep{ 2008PhRvL.100i1303C}. \hi\ IM can be conducted via varying instrumental set-ups, usually designed to derive greatest sensitivities on the largest scales $>1$ deg with a large collecting area for instantaneous high sensitivities. Examples of such telescopes include single-dish telescopes, dish interferometers, cylindrical dish interferometers, and aperture arrays. 

The most commonly used observing strategy for cosmological \hi\ IM at low redshifts is the use of single-dish telescopes with a field-of-view varying between arcmins and degree scales which scan the skies by either drift mode or fast scanning. \hi\ IM experiments have been conducted at most single-dish telescopes with suitable frequency coverage and radio quiet environment and most previous measurements are considered as SKAO pathfinder experiments, including the Green Bank telescope \citep{2010Natur.466..463C, 2013ApJ...763L..20M, 2013MNRAS.434L..46S, 2022MNRAS.510.3495W}, Murriyang (Parkes) Telescope \citep{2018MNRAS.476.3382A, Li_2021}, and FAST telescope \citep{2020MNRAS.493.5854H, 2023ApJ...954..139L, 2025ApJS..279...32Y}. Dish interferometers for \hi\ IM circumvent the technical challenges of a large single-dish and - if purpose built- are usually designed with a dense core for increased sensitivities for the largest scales. For example, the Indian Giant Metrewave Radio Telescope (GMRT, see e.g. \citealt{2011MNRAS.411.2426G, Chakraborty2021, Elahi2024}) consisting of 30 fully steerable 45m dishes has been conducting several \hi\ IM campaigns at intermediate redshifts $z\sim2.2$. The HIRAX experiment is currently under construction with the ultimate goals to operate 1024 6m dishes targeting $0.78<z<2.55$ \citep{2016SPIE.9906E..5XN, Crichton:2021hlc}. SKAO pathfinder experiments in the context of \hi\ IM are described in \cite{Elahi01.2026.SKA}. 

Arguably one the most pertinent experiments in the build-up to the SKAO is the single-dish \hi\ IM survey conducted with MeerKAT, the MeerKAT Large Area Synoptic  (MeerKLASS\footnote{\url{https://meerklass.org/}}, \citealt{MeerKLASS:2017vgf}). In this experimental set-up, each dish of the array simultaneously observes the same sky area and the auto-correlations of the recorded visibilities are used to create sky maps \citep{2015aska.confE..19S}. Within MeerKLASS, a fast scanning approach is employed in lieu of a drift scan which fully leverages the gain stabilities on a 90 min intervals to scan 300 square degree sky area. The fast scans are done in constant elevation to minimise atmospheric and ground spill fluctuations and each sky area is scanned up to 30 times \citep{Wang:2020lkn}. A monumental benefit of this survey is the commensal creation of high-resolution maps from the visibility data via On-the-fly mapping approaches \citep{PaulOTF, ChatterjeeOTF, ManglaOTF}. Even though sensitivities are too low to target the cosmological \hi\ signal, there is a wealth of astrophysics as well as cosmology information within the data which can be obtained from continuum maps and polarisation cubes. An intriguing aspect is the prospect of transient detection within the repeated blocks of observations, an overview is provided in \cite{Chatterjee01.2026.SKA}. 

As MeerKAT currently contains about one third of dishes compared to the full Baseline Design of SKA-Mid, the MeerKLASS experiment is highly indicative to the SKA capabilities as usual considerations of scalability and resolution between MeerKAT and SKA are much diminished for the single-dish case. MeerKLASS has successfully conducted several pilot surveys with signal detections at $z \sim 0.4$ \citep{2023MNRAS.518.6262C, MeerKLASS:2024ypg, Carucci:2024qpm} and is currently surveying the Southern Skies at $0.5<z<1.5$ aiming at a BAO detection. A full overview of MeerKLASS in the context of SKAO can be found in chapter \cite{Cunnington01.2026.SKA}.

MeerKAT observations have also been used to derive the first interferometric detection of the \hi\ IM signal at redshift, $z\sim0.4$ \citep{Paul:2023yrr} using deep tracking observations, with a follow-up upper limits derived from mosaicked MIGHTEE observations \citep{2025MNRAS.541..476M}. The signal obtained from MeerKAT visibilities contains mostly information on cosmic \hi\ evolution and its relation to dark matter haloes. However, it should be noted that for higher redshifts $z>2$, visibility data from the SKA-Mid might be the only avenue for BAO constraints from the SKAO in the near future, an overview is provided in \cite{Mazumder01.2026.SKA}.

Generally, synergies between \hi\ IM surveys with the SKAO and more traditional observational campaigns for cosmology at optical/near-infrared frequencies have been shown to be potentially transformational and have drawn significant interest by the community over the past decade, since the previous SKA Science Book \citep[see][for cosmology]{Maartens:2015mra}. The panorama of such cosmological surveys is vast, but the state of the art is certainly set by the Dark Energy Spectroscopic Instrument \citep[DESI,][]{2022AJ....164..207D,2025arXiv250314745D}, the Legacy Survey of Space and Time at the Vera C.\ Rubin Observatory \citep[LSST,][]{2019ApJ...873..111I,2018arXiv180901669T}, and the European Space Agency’s flagship, the \textit{Euclid} satellite \citep{laureijs_2011,2025A&A...697A...1E}. In addition to so-called Stage V galaxy redshift surveys envisaged for the coming decades (like MegaMapper, \citealt{2022arXiv220904322S}, or the Wide-field Spectroscopic Telescope, \citealt{2024arXiv240305398M}), which will allow for transformational synergies \citep[see e.g.][]{barberi2024}.

If the aforementioned experiments/facilities in the optical and near-infrared bands arguably exhaust the state of the art for the study of the cosmic LSS, the cosmology community is putting a significant effort at lower frequencies as well. A new generation of telescopes targeting the cosmic microwave background (CMB) is taking over from the highly successful campaigns of the last decade, most notably the Planck satellite \citep{Planck:2018vyg}, the Atacama Cosmology Telescope \citep{2025JCAP...11..062L}, and the South Pole Telescope \citep{2026PhRvD.113h3504C}. Among them, the Simons Observatory \citep{2025JCAP...08..034A} will deliver an order-of-magnitude improvement in sensitivity and mapping speed over previous CMB experiments, enabling high-resolution, low-noise measurements of temperature and polarisation anisotropies across a broad range of angular scales. In the further future, Stage-4 CMB \citep{2016arXiv161002743A} observations will provide a transformative leap, combining unprecedented sensitivity, angular resolution, and sky coverage to enable next-generation constraints on inflation, neutrino physics, dark matter, and the growth of cosmic structure.

The compiled science case for cosmology with the SKAO has been presented in \cite{SKA:2018ckk}. In this chapter, we follow the proposed survey and science cases of previous work in \cite{SKA:2018ckk}, and provide updated SKAO predictions for $0<z<6$. SKA-Mid will be limited to $z<3$ for \hi\ IM, however, SKA-Low provides further capabilities to conduct \hi\ IM surveys for $3<z<6$. These experiments would be conducted commensally with EoR observations and cosmology forecasts using the higher frequencies are presented within this chapter. 

This chapter reviews the baseline \hi\ IM capabilities of the SKAO to constrain $\Lambda$CDM cosmology for $0<z<6$. However, owing to the immense advancement in theory, pipeline and data analysis methodology, there are several chapters dedicated to reviewing progress in individual areas, e.g. methodology in \cite{Spinelli01.2026.SKA}, simulations and models in \cite{Ronconi01.2026.SKA}, and cosmology with higher-order statistics in \cite{Majumdar01.2026.SKA}.

\section{SKA \hi\ Intensity Mapping Survey}

\subsection{Observational parameters}
This chapter focuses on baseline predictions for cosmology with \hi\ IM given a set of standard survey definitions. As previously mentioned following \cite{SKA:2018ckk}, we focus on single-dish IM with SKA-Mid and commensal SKA-Low observations for this chapter, defined as follows.
\begin{itemize}
\item \wideims~: SKA-Mid AA4 in Band 1 with redshift range $z=0.35-3$ covering $20,000\,{\rm deg}^2$ and an integration time of $t_{\rm tot}= 10,000$ hrs on sky. 
\item \deeplowims~: SKA-Low AA4 with $100 {\rm \, deg}^2$ sky coverage and an integration time of approximately $t_{\rm tot}= 5,000$ hrs on sky using data from sub-bands at frequencies $200-350\rm MHz$, equivalent to $3<z<6$.
\end{itemize}
The following forecasts are based on Fisher forecasts, formalism outline in \autoref{sec:methods}. Unless otherwise specified, we do not include any errors introduced by observational systematics in the prediction as the goal of this chapter is to determine the best case constraining power of the \hi\ Intensity Mapping with the SKAO. Observational challenges are discussed in the following section, and more details on the realities of large \hi\ IM surveys including `Lessons learned' can be found in the pathfinder and precursor chapters \cite{Elahi01.2026.SKA}.

\subsection{Observational challenges}
\label{sec:obschallenges}

The \hi\ signal is inherently faint, and sub-mK sensitivity is required per volume element to detect the 21 cm brightness temperature fluctuations directly, although statistical detections of the power spectrum are possible with higher noise levels. Beyond raw sensitivity, the main challenges surrounding 21 cm cosmology with SKAO are due to systematic contamination from instrumental effects and bright foreground emission.

\subsection{Foreground separation}
Foreground contamination from Galactic synchrotron and free-free emission, plus similar extragalactic components, is around four orders of magnitude brighter than the target 21 cm signal away from the Galactic plane. The foreground emission spectra are spectrally smooth, and the emission is greatest on large angular scales, providing a useful distinction from the rapidly spectrally-varying 21 cm signal that can be used to separate it from these other components. The 21 cm signal does have smooth (low-$k_\parallel$) radial modes however, which are difficult to unambiguously distinguish from foregrounds, while leakage of polarised foregrounds can exhibit significant spectral variability. The loss of these larger-scale modes of the 21 cm field can generally be tolerated as long as the scales of interest (e.g. the BAO scale) are relatively unaffected.

The foreground emission is not perfectly reconstructed by the instrument however; in particular, the chromaticity and angular structure of the primary beam modulates the otherwise spectrally-smooth emission, introducing spurious spectral structure that can obscure a significant fraction of the Fourier space spanned by the observations. Without very accurate models of both the foreground emission and instrumental response, it is difficult to cleanly subtract this contamination. One approach, common for interferometric experiments, is to excise all of the data in the contaminated `foreground wedge' or `pitchfork' region of Fourier space. This significantly reduces the number of modes available for measuring the 21 cm signal, but removes the need to make detailed models of the sky and instrument. Great care must be taken to avoid additional leakage of modes from within this `foreground avoidance' region into the rest of the Fourier space, e.g. through internal reflection effects and ringing due to missing data. 

Another approach is to use data-driven (or `blind') foreground removal methods that seek to learn the spectral structure of the data, e.g. through principal component analysis, and subtract out the modes with the largest variance \citep{Wolz:2013wna, Alonso:2014dhk, 2015ApJ...815...51S, Cunnington:2020njn, Spinelli:2021emp}. This has been most successfully applied to auto-correlation observations, e.g. with GBT, Parkes, and MeerKAT. Increasing numbers of modes must be subtracted for more complex instrumental responses, which also has the effect of removing a substantial fraction of the 21 cm signal itself. This `signal loss' must be corrected for; this is typically done by injecting mock \hi\ signals and cleaning those along with the real data to estimate a `transfer function' that can then be applied to recover a \hi\ power spectrum that has been corrected for signal loss \citep{2015ApJ...815...51S, Cunnington:2023jpq}.

Some approaches have trialled the use of Machine Learning or Bayesian forward modeling using Bayesian methods for the foreground separation, though generally methods have not been advanced enough for data application.  

\subsection {Calibration}
Uncertainties in the foreground and instrument models also present challenges for calibration of the data. 
A variety of methods are used to calibrate these components, depending on whether interferometric or autocorrelation observations are being made. For instance, autocorrelation measurements are affected by gain variations that drift in time in a stochastic way -- often called $1/f$ noise. This can be mitigated by scanning the telescope quickly across the sky, so angular separations of interest are covered before the gain drift becomes significant \citep{Li:2020bcr, Irfan:2023njr}. Internal calibration references can also be used to correct for the gain drifts. Filtering of the data time series can also be used to suppress this effect, particularly if the gain drifts are strongly correlated in frequency. Other calibration parameters, such as the overall flux scale, bandpass, pointing error can be calibrated by observing bright astrophysical sources, which must be modelled. Contributions to the system temperature due to the atmosphere and ground are generally also accounted for via modelling. Several existing pipelines are capable of performing most of the required calibration tasks, and have been used on MeerKLASS data, namely KATcali presented in \cite{Wang:2020lkn} and MuSEEK. 

Interferometric observations correlate out $1/f$ noise, but must still determine the per-antenna receiver gains as a function of time and frequency. Some radio arrays are designed with a high level of redundancy (i.e. many baselines of the same length and orientation) to aid calibration, but for the SKAO arrays, sky-based calibration will generally be required. This effectively compares the observations at any given point in time with a detailed sky model, perhaps around a bright source, and assumes that any differences between the model and observations can be attributed to multiplicative gains. For high dynamic range, wide-field observations of the kind the SKAO arrays are capable of, this is likely to require a large and detailed sky model, and may well need to be iteratively constructed from SKAO observations, as existing source catalogues etc. are not sufficiently complete. Sky model errors can also cause part of the signal to be erroneously absorbed into the calibration solutions, resulting in additional spectral and temporal structure. Smoothing the gain solutions can help reduce the severity of the gain errors.

\subsection{Specific challenges for the SKAO}
Some specific calibration challenges have been observed with SKAO precursors that may also be applicable to the SKAO arrays. For SKAO-MID, one is the `zebra stripes' observed in MeerKLASS survey data, caused by a time-dependent gain non-linearity effect as the dishes scan past a bright out-of-band transmitter on the horizon. This leads to large ripple structures in the observed maps. 

Another involves the chromaticity of the primary beam; low-level ripples are observed in the beam width as a function of frequency, which appear to be associated with standing waves set up between the edges of the metal plates used to build the dish surface and the feed \citep{Matshawule:2020fjz}.
This frequency structure of the beam will convolve with any input signal, generating extra temperature fluctuations along frequency. Such fluctuations are small and will not impact the \hi~signal directly but can be problematic when convolved with strong temperature contaminants that are usually smooth in frequency. This is particularly true for the ground pickup and sky foregrounds. Ground pickup can be dealt with in the time domain by keeping the dish elevation constant. Convolution with sky foregrounds, such as the hot galactic synchrotron, has been shown to affect foreground removal and introduce mode coupling in the \hi~power spectrum.
There will also be the additional complication of multiple beam shapes caused by the use of a heterogeneous array made of MeerKAT, MeerKAT$+$ and SKA-Mid dishes. While this can in principle be dealt with at the mapmaking level, either through beam deconvolution or simple re-smoothing to a common low resolution, this is yet to be explored in practice. A good knowledge of the beam through holographic measurements will be crucial for this process \citep{2023AJ....165...78D}.

For SKAO-LOW, the mutual coupling artifacts -- shifted copies of the sky signal caused by multiple signal paths being combined in the correlator -- have also affected precursors such as HERA, and have the potential to be a limiting systematic if they cannot be modelled accurately enough.

\section{Models and methodology}
\label{sec:methods}
The primary aim of \hi\ Intensity Mapping is detecting the \hi\ signal via the power spectrum $P_\hi(k,\mu, z)$ acting as a tracer for the matter power spectrum $P_{\rm m}(k,z)$. $P_\hi(k,\mu)$ can be modelled as 
\begin{equation}
\begin{split}
    P_\hi(k,\mu, z) = \overline{T}^2_\hi(z) & \Big(b_\hi(z) + f(z) \mu^2\Big)^2 P_{\rm m}(k,z) \\
\end{split}
\end{equation}
where $b_\hi$ is the \hi\ bias, $\overline{T}^2_\hi$ the averaged \hi\ brightness temperature, and $f$ the growth factor. The power spectrum is a function of wavenumber $k$ and $\mu$ which is the cosine $k_\parallel/k$ between the wave-vector and the line of sight. the \hi\ temperature and the \hi\ bias are degenerate and constraints on individual parameters can only be achieved by obtaining directional constraints of redshift space distortions. 

The overall amplitude of the power spectrum scales with the mean \hi\ brightness temperature which can be calculated via
\begin{equation}
    \overline{T}_\hi(z) = 180\, \Omega_\hi\frac{ h \, (1+z)^2}{ H(z)/H_0}\,\text{mK}\,,
\end{equation}
where $\Omega_\hi$ is the global density of the \hi\ gas. $\Omega_\hi$ can be modelled via 
\begin{equation}
    \Omega_\hi(z) = 0.00067432 + 0.00039\,z - 0.000065\,z^2\,.
\end{equation}

which is based on the fitting function in \citet{SKA:2018ckk} but incorporate results from MeerKLASS intensity mapping \citep{Cunnington:2022uzo}.

The \hi\ bias is usually modelled with an empirical model, for example using hydrodynamical simulations \citep{Villaescusa-Navarro:2018vsg} which leads to
\begin{equation}
    b_\hi(z) = 0.842 + 0.693\,z - 0.0459\,z^2\,.
\end{equation}
Note that this base theoretical model can be extended to incorporate the Alcock–Paczynski effect, modulations of Primordial Non-Gaussianity and other effects.  

In order to predict survey constraints, basic observational effects such as beam convolution and additive instrument noise need to be modelled. The instrument noise for auto-correlations can be written as 
\begin{equation}\label{eq:sigma_N}
    \sigma_{\mathrm{N}}(\nu)=\frac{3\,T_{\mathrm{sys}}(\nu)}{\sqrt{2\, \delta\nu \, N_{\text{dish}} \,t_{\text{tot }}\,\theta_{\mathrm{FWHM}}^2 /  A_{\text{sur}}}}\,.
\end{equation}
where $T_{\mathrm{sys}}$ is the system temperature of the telescope, $\delta \nu$ the frequency width of the maps, $N_{\text{dish}}$ the number of dishes used in the observation, $t_{\rm tot}$ the total on sky survey time, $\theta_{\mathrm{FWHM}}$ the full-width half maximum of the dish beam, and $A_{\rm sur}$ the total survey area. The noise contribution in the power spectrum space can be calculated via $P_\text{N}=V_\text{pix}\sigma_\text{N}^2$, where $V_\text{pix}$ is the comoving volume of one pixel. The resulting power spectrum is 
\begin{equation}\label{eq:P_HI_obs}
    P_\hi(k,\mu,z) = P_\hi(k,\mu,z)\,\mathcal{B}^2_{\rm beam}(k,\mu,z) + P_{\rm N}(z)\,.
\end{equation}
For our predictions, the telescope beam is assumed a Gaussian function with uniform width, i.e. effects of differing dishes are neglected. 

The fiducial cosmological model used is \LCDM, with parameters determined by the 2015 Planck analysis (TTTEEE+lowP) \citep{Planck:2015fie}. The values used are the same as in \citet{SKA:2018ckk} and are listed for reference in Table \ref{table:cosmo_fid}.

\begin{table}[t]
 \centering
 \begin{tabular}{c c c c c c c}
  $\Omega_{\rm b}\,h^2$ & $\Omega_{\rm c}\,h^2$ & $h$ & $\ln(10^{10}\,A_{\rm s})$ & $n_{\rm s}$ & $\tau$ & $\Sigma m_\nu$ [eV]  \\ \hline\hline
    $0.02225$ & $0.1198$ & $0.6727$ & $3.094$ & $0.9645$ & $0.079$ & $0.06$
 \end{tabular}
 \caption{The fiducial values used for the flat $\Lambda$CDM cosmology, as measured by Planck 2015 \citep{Planck:2015fie}. }
 \label{table:cosmo_fid}
\end{table}

\section{\hi\ power spectrum }

In the following, we investigate the constraining power on the standard model of cosmology from measurements of the \hi\ auto power spectrum, $P_{\rm HI}(k,\mu,z)$, for the \wideims. Following the methodology developed in~\cite{Berti:2022ilk,Berti:2023viz,Autieri:2025sxz}, we construct synthetic data sets of $P_{\rm HI}(k,\mu,z)$ observations and conduct a Bayesian analysis to constrain the cosmological parameters from SKAO \hi\ intensity mapping measurements alone and combined with CMB.\footnote{The likelihood code used for this analysis is available at \url{https://github.com/mberti94/topk}.} In this section, we assume a Planck 2018 fiducial cosmology, and we model the signal and uncertainties to mimic SKA measurements as described above. In the modelling for the \hi\ power spectrum we include non-linearities and Alcock-Paczynski effects. From the synthetic data set for the $P_{\rm HI}(k,\mu,z)$ monopole and quadrupole at six equispaced effective redshifts in the range $z\in [0,3]$, we constrain the cosmological parameters $\{\Omega_b h ^2, \Omega_c h^2, H_0, \tau, n_s, \ln(A_s)\}$. Along with the parameters describing the $\Lambda$CDM model, for each redshift bin we vary a set of nuisances relative to the \hi\ likelihood function, i.e. the shot noise and combinations of the brightness temperature, the \hi\ bias, and the growth rate. We study two different cases, one in which we assume some prior knowledge on the brightness temperature and the \hi\ bias can be inferred by other observations (labelled as \textit{optimistic}) and one in which we vary all the nuisances imposing wide flat priors (labelled as \textit{pessimistic}).

\begin{figure}
\centering
\includegraphics[width=0.49\textwidth]{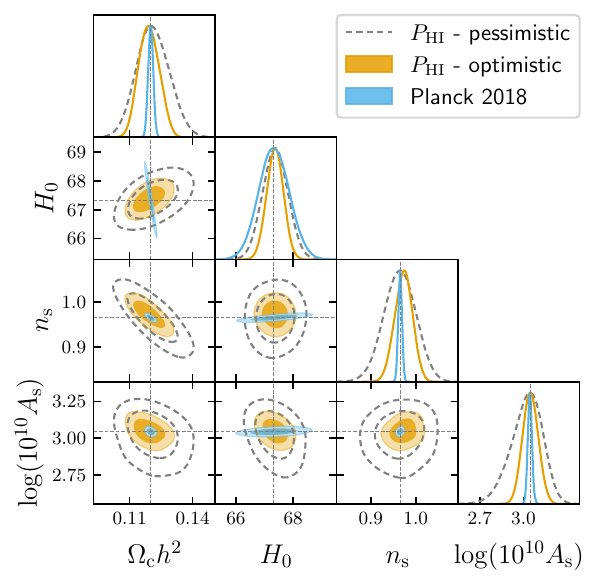}
\includegraphics[width=0.49\textwidth]{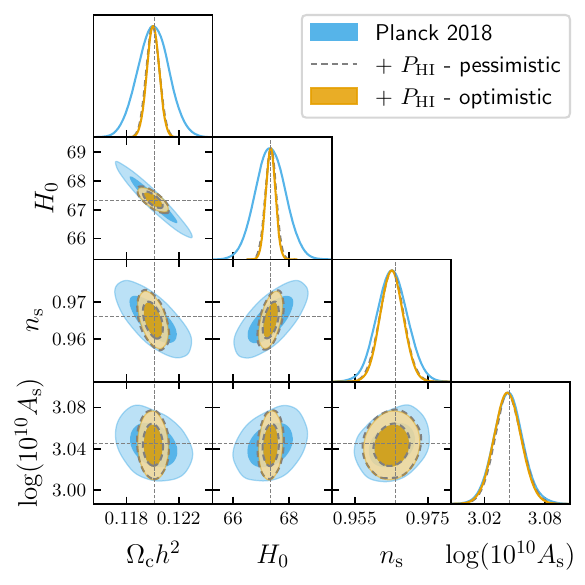}
\caption{\label{fig:HI_power_spectrum_forecasts} Forecasted joint constraints (68\% and 95\% confidence regions) and marginalized posterior distributions on cosmological parameters from the \hi\ power spectrum alone (left panel) or in combination with Planck data (right panel). Here, the label "Planck 2018" stands for the  TT, TE, EE + lowE + lensing likelihoods and data sets~\cite{Planck:2018vyg,planck:2018like,planck:2018_maps}. Dashed gray vertical lines mark the fiducial cosmology used to construct the $P_{\rm HI}$ synthetic data sets. Figure adapted from \citet{Berti:2023viz}.}
\end{figure}

The forecasted constraints from this analysis are shown in Figure \ref{fig:HI_power_spectrum_forecasts}. Results using the SKAO \hi\ power spectrum synthetic data set alone (left panel) suggest that \hi\ intensity mapping could provide constraints competitive with state-of-the-art observations. In particular, in all the considered scenarios, we obtain a constraint on $H_0$ comparable with the estimate coming from CMB measurements. Already in the pessimistic scenario we find $H_0 = 67.39\pm 0.44$ km s$^{-1}$ Mpc$^{-1}$ to be compared with the Planck alone constraint $H_0 =  67.36 \pm 0.54$ km s$^{-1}$ Mpc$^{-1}$
~\citep{Planck:2018vyg}. In the optimistic case we are able to further reduce the error on $H_0$ to $0.29$ km s$^{-1}$ Mpc$^{-1}$. This is due to the fact that, in particular in the lowest bin, the \hi\ IM data set is sensitive to the growth of structure at non-linear scales, which is not well probed by CMB data and mostly aided by the BAO information that is modeled through the Alcock-Paczynski effect in the \hi\ IM data sets. Additionally, in both cases we observe a strong correlation in the $H_0 - \Omega_c h^2$ plane. This feature is ascribable to the dependence on the matter power spectrum. A measure of the \hi\ multipoles would fix the shape of the matter power spectrum that can be shown to be dependent on $\Omega_m h$. This implies that $\Omega_m h^2$, and consequently $\Omega_c h^2$, is correlated with $h$ and $H_0$. This correlation is pivotal when combining intensity mapping data with CMB measurements.  

We also test if and how \hi\ intensity mapping could improve the current constraints on $\Lambda$CDM when used in combination with other probes. We thus constrain the cosmological parameters using the synthetic data set we construct along with Planck 2018 CMB observations. We emphasize that, when the synthetic data are combined with real observations, we are primarily interested in the resulting error bars, while the resulting marginalized means serve primarily as a check that the method successfully recovers the input
fiducial cosmology. As shown in the right panel of Figure \ref{fig:HI_power_spectrum_forecasts}, adding the \hi\ power spectrum multipoles to the CMB, significantly increases the constraining power on $\Omega_c h^2$ and $H_0$. This gain is due to the combination of opposite correlation directions between the CMB and the \hi\ power spectrum on these cosmological parameters. The combination between the two significantly reduces the degeneracy, thus improving on the estimated error already within the pessimistic case. For example, we obtain $H_0 = 67.34\pm 0.17$, where the error is more than three times smaller than the result from Planck alone. 

\begin{figure}
\centering
\includegraphics[width=0.49\textwidth]{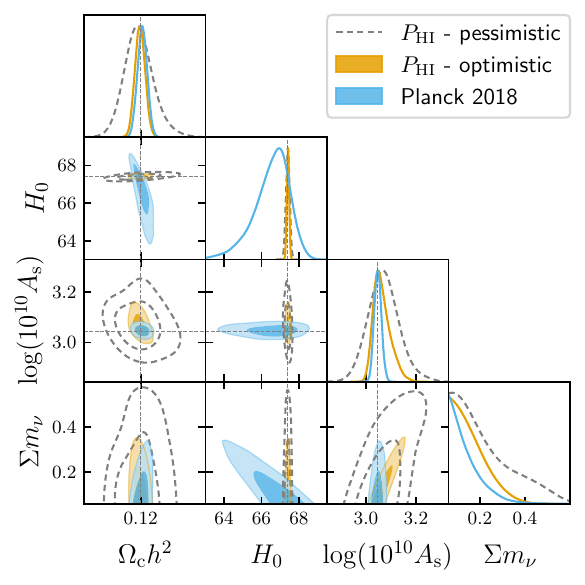}
\includegraphics[width=0.49\textwidth]{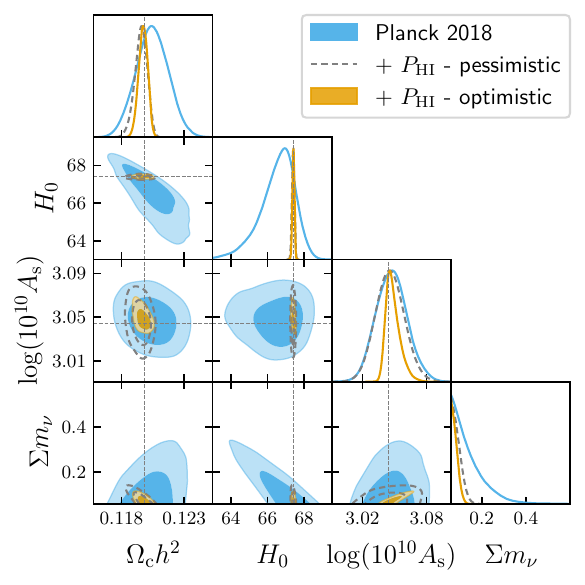}
\caption{\label{fig:HI_power_spectrum_forecasts_mnu} Forecasted joint constraints (68\% and 95\% confidence regions) and marginalized posterior distributions on cosmological parameters and the total neutrino mass from the \hi\ power spectrum alone (left panel) or in combination with Planck data (right panel). Here, the label "Planck 2018" stands for the  TT, TE, EE + lowE + lensing likelihoods and data sets~\cite{Planck:2018vyg,planck:2018like,planck:2018_maps}. Dashed gray vertical lines mark the fiducial cosmology used to construct the $P_{\rm HI}$ synthetic data sets. Figure adapted from \citet{Autieri:2025sxz}.}
\end{figure}

Measurements of the \hi\ power spectrum will also be instrumental in further improving the constraints on the neutrino mass. As presented in~\cite{Autieri:2025sxz}, we expand the methodology presented above to accurately model massive neutrino cosmology and forecast constraints on the total neutrino mass $\Sigma m_\nu$. We present the results of our analysis in Figure~\ref{fig:HI_power_spectrum_forecasts_mnu}. We find that the \hi\ synthetic data sets alone are able to provide upper limits on $\Sigma m_\nu$ comparable with single-instrument CMB observations. At the 95\% confidence level, we obtain $\Sigma m_\nu < 0.287$ eV in the optimistic case and  $\Sigma m_\nu < 0.425$ eV in the pessimistic case, to be compared with the Planck alone upper limit of $\Sigma m_\nu < 0.280 $ eV. We note that the more recent analysis obtained combining SPT, ACT, and Planck together \citet{SPT2025} provide rather more stringent results than IM alone, with an upper limit of $\Sigma m_\nu < 0.19 $ eV at the 95\% confidence level. When Planck 2018 data are added to the HI power spectrum, the constraints significantly improve with respect to HI alone results, similarly in both the considered cases, with $\Sigma m_\nu < 0.105$ (optimistic) and $\Sigma m_\nu < 0.126$ eV (pessimistic). 

We conclude that \hi\ auto power spectrum SKAO observations will provide a new, competitive cosmological probe, complementary to CMB and pivotal for gaining statistical significance on the cosmological parameters
constraints.

\subsection{Baryon Acoustic Scale}

As a tracer of the large-scale structure, \hi\ temperature fluctuations will include imprints of the baryon acoustic oscillations (BAO). The BAO feature can be used as a very robust probe of the expansion history of the Universe -and thus, dark energy- through the Alcock-Paczynski effect \citep{2006astro.ph..9591A}. In order to forecast the SKAO sensitivity, we adapt the template-based approach applied to galaxy surveys to the LIM case, shown to be also robust against beyond-$\Lambda$CDM physics at early times \citep{2020PhRvD.102l3515B}. We consider only \hi\ autocorrelations; sensitivity is expected to significantly improve when combined with galaxy clustering in a multi-tracer analysis. 

The template adopted for the \hi\ power spectrum Legendre multipoles is
\begin{equation}
    P_\ell(k) =\frac{2\,\ell+1}{2\,\alpha_{\rm iso}^3}\,\int_{-1}^{1}{\rm d}\mu\,\mathcal{A}\,\left[\frac{1+\beta(\mu')^2}{1+(k'\,\mu'\,\sigma_{\rm p})^2}\right]^2\,P_{\rm dw}(k',\mu')\,\mathcal{B}(k',\mu')\,\mathcal{L}_\ell(\mu)+\mathcal{D}_\ell(k)\,,
\end{equation}
where $\mathcal{A}$ is an overall amplitude encoding $\langle T_{\hi}b_{\hi}\sigma_8\rangle$, the term in brackets accounts for redshift space distortions at large and small scales, $P_{\rm dw}$ is the partially de-wiggled matter linear matter power spectrum which accounts for the smearing of the BAO due to bulk nonlinear clustering - computed assuming a fiducial cosmology and following the expression in \cite{2013MNRAS.430.2446W}, and depends on the characteristic velocity dispersion length parameter $\sigma_{\rm v}$, as adopted by the Euclid Consortium \citep{2025A&A...693A..58E}-, and $\mathcal{B}$ accounts for the beam suppression, which we assume to be Gaussian with full-width half maximum $\theta_{\rm FWHM}$. In the expression above, $k$ and $\mu$ are the \textit{observed} wavenumber and cosine $k_\parallel/k$ between the wave-vector and the line of sight, obtained after transforming redshifts and positions on the sky to a three-dimensional map. This step requires the assumption of a fiducial cosmology, and therefore the \textit{observed} quantities may differ from the \textit{true} ones, $k'$ and $\mu'$, if the assumed cosmology does not match the true one. Their relationship is given by \cite{1996MNRAS.282..877B}
\begin{equation}
    k'=\frac{k\,\alpha_{\rm AP}^{1/3}}{\alpha_{\rm iso}}\,\left[1+\mu^2\,\left(\frac{1}{\alpha_{\rm AP}^2}-1\right)\right]^{1/2}\,\qquad\qquad \mu'=\frac{\mu}{\alpha_{\rm AP}}\,\left[1+\mu^2\left(\frac{1}{\alpha_{\rm AP}^2}-1\right)\right]^{-1/2}\,.
\end{equation}
Here, $\alpha_{\rm iso}$ and $\alpha_{\rm AP}$ are the isotropic dilation and the anisotropic distortion in the distances, respectively, and are related to rescaling of scales along and transverse to the line of sight as
\begin{equation}
    \alpha_\parallel=\alpha_{\rm iso}\,\alpha_{\rm AP}^{2/3}=\frac{\left[H(z)r_{\rm d}\right]^{\rm fid}}{H(z)r_{\rm d}}\,,\qquad\qquad \alpha_\perp=\frac{\alpha_{\rm iso}}{\alpha_{\rm AP}^{1/3}} = \frac{D_M(z)\,r_{\rm d}}{\left[D_M(z)\,r_{\rm d}\right]^{\rm fid}}\,, 
\end{equation}
respectively, where $H$ is the Hubble parameter, $D_M$ is the comoving angular diameter distance, and $r_{\rm d}$ is the sound horizon at radiation drag, and we distinguish between the fiducial choices and the actual values.\footnote{We need to include the difference between the fiducial and actual $r_{\rm d}$, because it determines the position of the BAO in our template, and changes in its value are completely degenerate with $\alpha_{\rm iso}$.} We do the forecast in terms of $\alpha_{\rm iso}$ and $\alpha{\rm AP}$, which present lower degeneracies, and then transform to $\alpha_\parallel$ and $\alpha_{\rm perp}$. Finally, $\mathcal{D}_\ell$ is a set of piecewise cubic spines, with coefficients $\boldsymbol{a}$ different for each multipole and patch observed, that model potential deviations from the template in the broadband of the power spectrum multipoles, as implemented in DESI \citep{2024MNRAS.534..544C}. We do not consider any density field reconstruction to sharpen the BAO feature \citep{2007ApJ...664..675E}, since even if it has been adapted to LIM \citep{2017JCAP...09..012O}, its performance has not been tested in the presence of foregrounds. 

\begin{figure}
    \centering
    \includegraphics[width=0.6\textwidth]{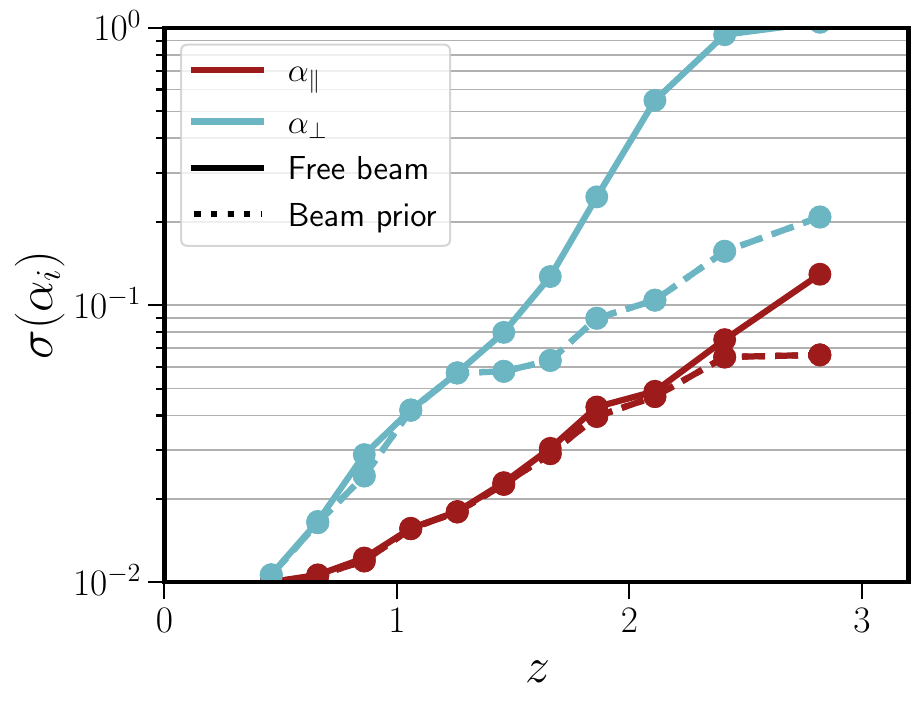}
    \caption{Marginalized forecast (relative) sensitivity of the SKAO-MID \hi\ wide survey to the Hubble parameter (red) and the comoving angular diameter distance (blue), expressed directly in terms of $\alpha_\parallel$ and $\alpha_\perp$, respectively. We show results without assuming any knowledge of the size of the beam (solid lines), and applying a Gaussian prior of 10\% uncertainty (dashed lines).}
    \label{fig:BAO}
\end{figure}

Then, our set of parameters for the forecast is $\lbrace \alpha_{\rm iso},\, \alpha_{\rm AP},\, \mathcal{A},\, \beta,\, \sigma_{\rm p},\,\sigma_{\rm v},\, \theta_{\rm FWHM},\, \boldsymbol{a}\rbrace$. We perform a different forecast for each redshift bin considered (as specified in \autoref{tab:bao}), marginalize analytically over $\boldsymbol{a}$, following \cite{2024MNRAS.534..544C}, and choose fiducial values \\
$\lbrace 1,\, 1,\, \langle T_{\hi}b_{\hi}\rangle^2,\, f/b_{\hi},\, 5\, {\rm Mpc},\, 5\, {\rm Mpc},\, 1.22c/(\nu_{\rm obs}D) \rbrace$, respectively. We consider a situation with no knowledge about $\theta_{\rm FWHM}$, and another one with a 10\% prior on it -which return similar results as no varying $\theta_{\rm FWHM}$ at all-. Forecast marginalized sensitivities are shown in Fig.~\ref{fig:BAO} and summarized in \autoref{tab:bao}.  

\begin{table}
\centering
\resizebox{\columnwidth}{!}{%
\begin{tabular}{lccccccccccc}
\toprule
$z_{\rm eff}$              & 0.45  & 0.65  & 0.85  & 1.05  & 1.25  & 1.45  & 1.65  & 1.85 & 2.10 & 2.40 & 2.80 \\
\midrule\midrule
$\sigma(\alpha_\parallel)$ & 0.010 & 0.011 & 0.012 & 0.016 & 0.018 & 0.023 & 0.029 & 0.040 & 0.047 & 0.065 & 0.067 \\
\midrule
$\sigma(\alpha_\perp)$     & 0.011 & 0.016 & 0.024 & 0.042 & 0.057  & 0.058  & 0.063  & 0.090 & 0.10 & 0.16 & 0.21 \\
\midrule
$\rho(\alpha_\parallel,\alpha_\perp)$ & $-0.49$ & $-0.49$ & $-0.43$ & $-0.46$ & $-0.40 $& $-0.29$ & $-0.20$ & $-0.18$ & $-0.16$ & $-0.11$ & 0.00 \\
\bottomrule

\end{tabular}%
}
\caption{Marginalized forecast (relative) sensitivity of the SKA-Mid \hi\ wide survey to the Hubble parameter and the comoving angular diameter distance, expressed directly in terms of $\alpha_\parallel$ and $\alpha_\perp$, respectively, using a 10\% uncertainty prior in the size of the beam. We also report the expected degeneracy/correlation between the parameters of interest at each redshift bin.}
\label{tab:bao}
\end{table}

The large telescope beam of the SKA-MID antennas may limit the potential of BAO analyses, as it sets a sensitivity limit in the determination of $D_M/r_{\rm d}$, especially at high redshift. Nonetheless, we also find that a very limited information of the beam (i.e., a 10\% prior information on the beam size) retains a large fraction of the information. In any case, we find that the measurements of $\alpha_\parallel$ will be competitive with DESI BAO measurements~\cite{DESI:2025zgx}. More importantly, the combination of both LIM and spectroscopic galaxy surveys will boost the overall BAO sensitivity, breaking parameter degeneracies with nuisance parameters and calibrating potential unknown systematic errors. These measurements will be essential to determine the nature of dark energy and whether it can be explain with a cosmological constant or it is indeed dynamical. 
\subsection{Turnover scale}

Alongside the detection of BAO in the \hi\ power spectrum (discussed in the previous section), another key large-scale feature is the \textit{turnover} in power on scales larger than the BAO. This turnover, usually denoted $k_0$, corresponds to the horizon size at matter–radiation equality and can be precisely calculated theoretically, making it a valuable standard ruler probe that is independent of the sound horizon. In this way, the turnover provides complementary information to BAO, helping to break degeneracies that affect sound-horizon-based measurements \cite[see e.g.][]{Farren:2021grl}. A first detection of this feature was reported with the WiggleZ survey \citep{Poole:2012ex}, and more recent galaxy surveys are beginning to improve these constraints \citep{Bahr-Kalus:2023ebd,Alonso:2024emk}. However, probing these ultra-large scales ($k\,{\lesssim}\,0.016\,h{\rm Mpc}^{-1}$) is extremely challenging for optical surveys due to limited volumes, while \hi\ intensity mapping at radio wavelengths is ideally suited to this task, offering rapid access to vast cosmological volumes.

\begin{table*}
    \setlength{\tabcolsep}{4.5pt}
	\centering
	\begin{tabular}{lccccccc}
		\toprule
		\textbf{\hi\ IM survey} & $z_{\min}$ & $z_{\max}$ & Area [$\text{deg}^2$] & \textcolor{black}{Volume [$({\rm Gpc}/h)^3$]} & $t_\text{obs}$ [hrs] & \textcolor{black}{$P_0/P_{\rm N}$} & $\alpha{>}0$ \\
		\midrule
        SKA-Mid Band 1 & 0.35 & 3 & 20,000 & \textcolor{black}{221.6} & 10,000 & \textcolor{black}{7.3} & 13.1$\sigma$\\		\toprule
		\textbf{Galaxy survey} &  &  &  &  & $N_\text{gal}$ &  & \\
		\midrule
        Stage III spectro-$z$ & 0.6 & 1.1 & 4,000 & \textcolor{black}{4.7} & $500{\times}10^3$ & \textcolor{black}{1.9} &  0.87$\sigma$\\
        Stage III photo-$z$ & 0.2 & 1.05 & 5,000 & \textcolor{black}{7.0} & $200{\times}10^6$ & \textcolor{black}{700.3} & 2.3$\sigma$\\
        Stage IV spectro-$z$ & 0.4 & 1.6 & 14,000 & \textcolor{black}{46.8} & $21{\times}10^6$ & \textcolor{black}{7.2} & 5.9$\sigma$\\
        Stage IV photo-$z$ & 0.3 & 3 & 20,000 & \textcolor{black}{225.6} & $10{\times}10^9$ & \textcolor{black}{445.9} & 16.9$\sigma$\\
        \bottomrule
	\end{tabular}
    \caption{Statistical significance of turnover detections ($\alpha{>}0$) for intensity mapping with the SKAO and comparison with hypothetical galaxy surveys in optical and near-infrared wavelengths. No consideration has been given to the impact from redshift uncertainties in the photo-$z$ surveys.}
    \label{tab:OpticalComparison}
\end{table*}

The turnover in the \hi\ power spectrum, $P_\hi$, can be constrained by using a model-independent fitting approach \cite[following][]{Poole:2012ex} in which the power spectrum around the turnover is fit with a piecewise parabolic model:
\begin{equation}\label{eq:ParabolaFit}
    \log _{10}\left(\frac{P_\hi(k)}{[{\rm mK}^2\,h^{-3}\,{\rm Mpc}^3]}\right)=\left\{\begin{array}{ll}
    \log _{10}(P_{0})\,\left(1-\alpha \,x^{2}\right) & \quad k<k_{0} \\
    \log _{10}(P_{0})\,\left(1-\beta \,x^{2}\right) & \quad k \geq k_{0}\,,
    \end{array}\right.
\end{equation}
where
\begin{equation}
    x = \frac{\log_{10}(k/[h/\text{Mpc}]) - \log_{10}(k_0/[h/\text{Mpc}])}{\log_{10}(k_0/[h/\text{Mpc}])}\,.
\end{equation}
This fit estimates the turnover scale $k_0$ together with three additional parameters. $P_0$ is the peak amplitude of the power spectrum, while $\alpha$ and $\beta$ govern the parabolic decline of power on either side of the turnover. Importantly, $\alpha$ is not simply a nuisance parameter: if confidence intervals favour $\alpha\,{>}\,0$, this constitutes statistical evidence for the presence of a turnover. Forecasts of the $\alpha$ constraint from SKA-Mid Band 1 intensity mapping, compared with those from optical galaxy surveys, are shown in \autoref{tab:OpticalComparison} \cite[see][for more details]{Cunnington:2022ryj}.

The recovered turnover scale $k_0$ can then be translated into a standard ruler, since its physical scale is set by the horizon size at matter–radiation equality. By mapping this to a dilation parameter, one can obtain a distance measurement that is directly sensitive to the Hubble constant, $H_0$. Forecasts indicate that SKA-Mid intensity mapping could achieve a ${\sim}\,2\%$ constraint on $H_0$ from the turnover feature alone \citep{Cunnington:2022ryj}.

\subsection{Ultra-large scales}

 As \hi\ IM can easily probe very large areas of the sky in a timely manner, it is very well suited to study physical phenomena on very large cosmological scales. Such science cases are discussed on their own in the SKAO science chapters: \citealp{Camera01.2026.SKA} for tests of gravity and \citealp{Fonseca01.2026.SKA} for ways to probe the very early universe.

\subsection{Synergies}

The scientific potential of \hi\ IM is greatly enhanced when one correlates it with other datasets. On one hand we can break degeneracies between cosmology and galactic astrophysics but we are also less sensitive to systematics as we expected them to be uncorrelated between different datasets. The simplest example is the one corresponding to the first detection of the \hi\ cosmological signal, cross-correlation between \hi\ IM and spectroscopic galaxy surveys \citealt{Chang:2007xk}, which was followed by more cross-correlation detections e.g. in \cite{2013ApJ...763L..20M, 2022MNRAS.510.3495W, 2023MNRAS.518.6262C}. In addition, cross-correlation between \hi\ and photometric galaxy survey can be used to probe the redshift distribution of the galaxy sample, giving a novel clustering redshift estimate \citep{2019MNRAS.482.3341C}.

The more elaborate synergies is the so-called multi-tracer technique \citep{Seljak:2008xr} which corresponds to joint inference and has a game changing potential in bias-like parameters. 

There has been an extensive literature looking at synergies between \hi\ IM and spectroscopic and photometric galaxy surveys, CMB, gravitational wave observatories and LIM with other lines.
Here, we showcase one example of synergies between SKA-Mid \hi\ IM and DES. For other examples, we refer the reader to \cite{SKA:2018ckk} as well as dedicated chapters in this Science Book, i.e. \citep{Baker01.2026.SKA, Fonseca01.2026.SKA}.



We perform a forecast for SKA-Mid in cross-correlation with Dark Energy Survey (DES), focusing on the constraints on the bias parameters of the tracers and the redshift scatter of the galaxy sample. We simulate log-normal mock observations in RA = (13,83) and Dec= (-55.5,-18), corresponding to the potential overlap between the SKA-Mid Wide Band 1 Survey and DES surveys. The galaxy catalogue is generated according to the redshift distribution $dN/dz$ and the photometric redshift uncertainty $\sigma_z$ derived from the DES Year~3 (Y3) Meta-Calibration sample \citep{2021MNRAS.505.4249M} in a simulation box with a sky area of $2100\, {\rm deg}^2$ ranging from redshift $z=0.4-1.1$, corresponding to the third tomographic bin with an effective redshift of $z_{\mathrm{eff}} = 0.74$. The photometric redshift uncertainty is approximated as $\sigma_z = 0.02$, consistent with the DES \textsc{RedMaGiC} characterization \citep{2021PhRvD.103d3503P}. We model the power spectra for the galaxy, the \hi, and their cross following Appendix A of \cite{2022MNRAS.516.5454R}.
We choose a fiducial cosmology model as reported in \cite{2020A&A...641A...6P} with the \hi\ bias $b_{\rm \hi} = 0.9$, the galaxy bias $b_g=1.5$, and the redshift scatter $\sigma_z=0.02$.
We generate 100 realizations of the mock observations, and average the 3D power spectrum into power spectrum monopoles for calculating the mean and the covariance of the signal. Since we are interested mostly in the redshift scatter, we only consider the galaxy auto and the cross-power spectrum.

Using this simulation pipeline, we then vary the fiducial parameters to perform Fisher matrix forecast (see e.g. Eq. (43) of \citealt{2020A&A...642A.191E}). We choose a step size of $1\%$ for calculating the partial derivatives.

\begin{figure}[htbp]
    \centering
    \includegraphics[width=0.7\textwidth]{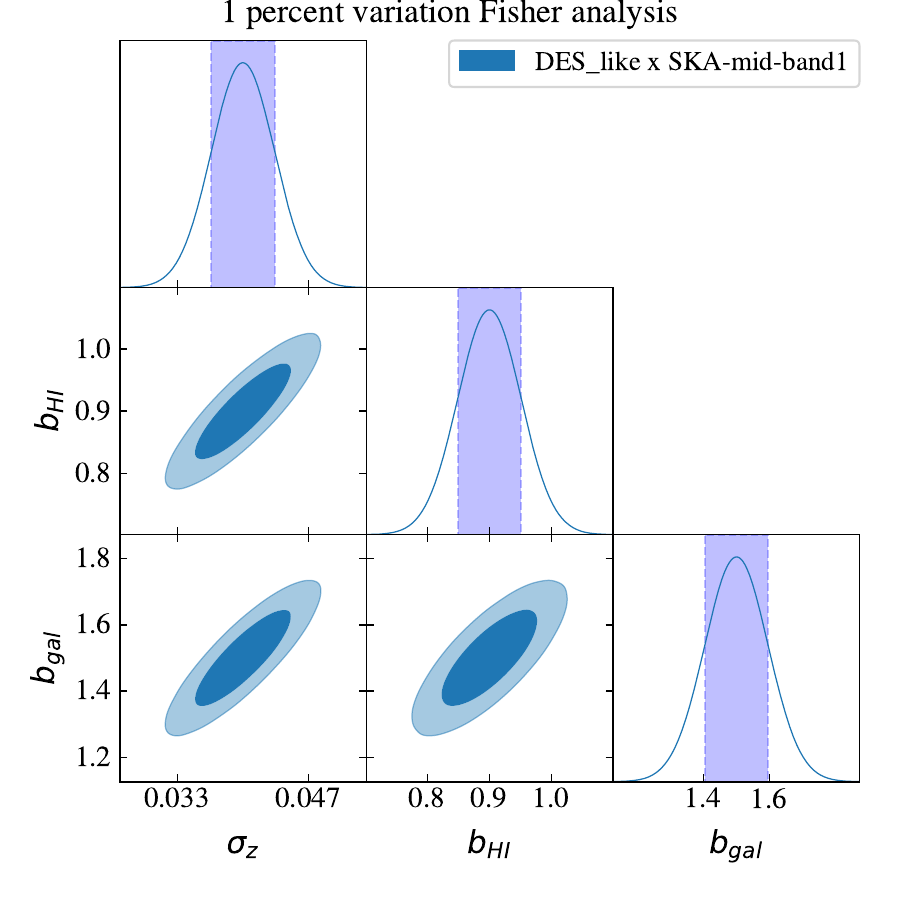}
    \caption{
        Corner plot showing the forecasted 1$\sigma$ and 2$\sigma$ confidence regions for the parameters 
        $\{b_{\rm HI},\,b_g,\,\sigma_z\}$ from the joint DES and SKA-Mid Wide Band 1 analysis. The diagonal panels show the marginalized 1D distributions, while the off-diagonal panels show the 2D parameter covariances. 
    }
    \label{fig:photozcross}
\end{figure}

The results are shown in \autoref{fig:photozcross}.
The forecasts show that for a cross-correlation analysis between DES and SKA-Mid, we will be able to achieve the constraints of $b_{\rm H\textsc{I}} = 1.5 \pm 0.09$, $b_g = 0.9 \pm 0.05$, $\sigma_z = 0.02 \pm 0.003$, resulting in a $\sim 8\%$ measurement of the redshift scatter of the galaxy sample. Comparing with the case using only the galaxy auto-power spectrum, we find that the constraints improve by a factor of $\sim 2.5$.

\subsection{Bi-spectrum}

\begin{figure}
\centering
\includegraphics[width=0.49\textwidth]{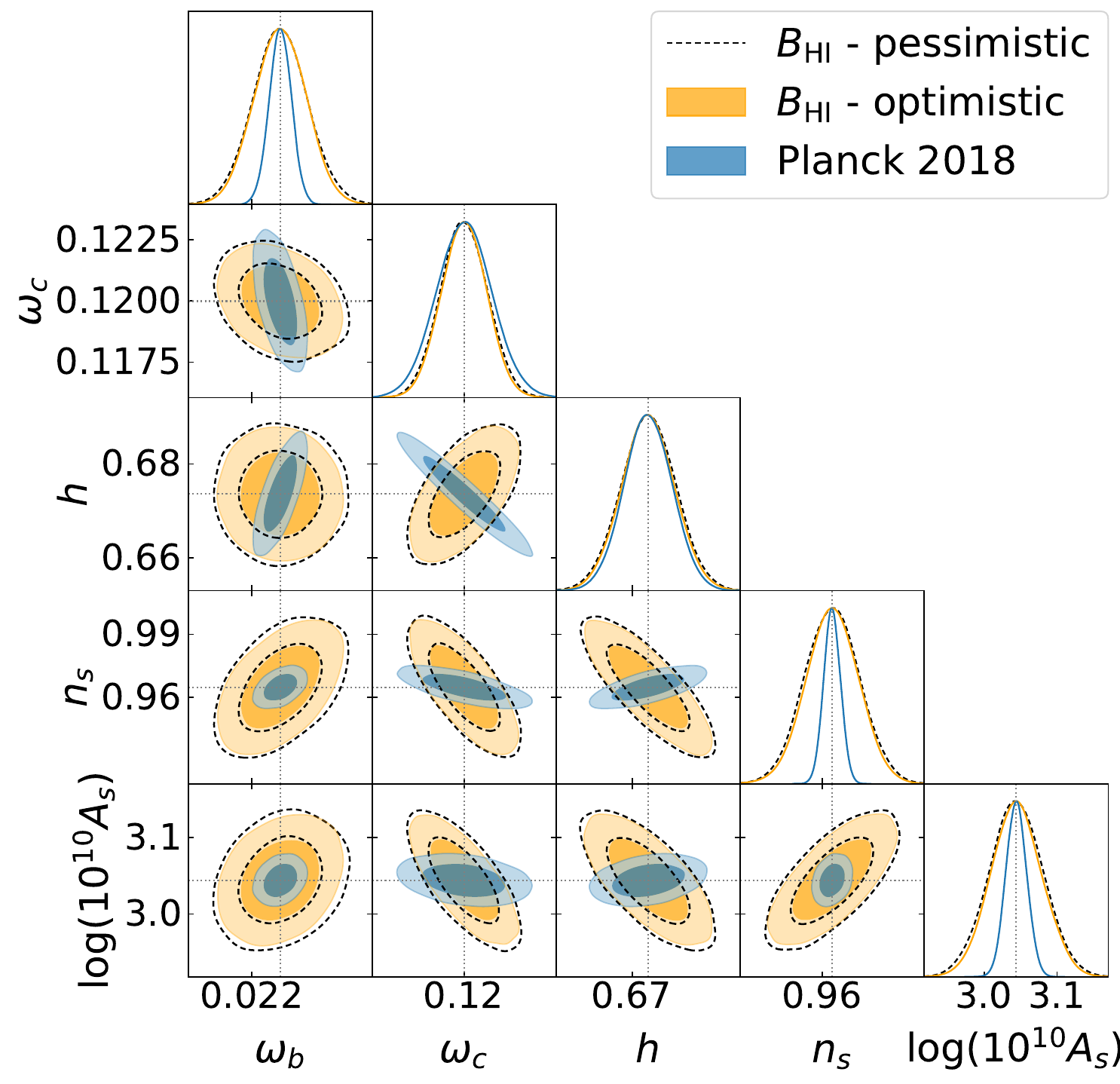}
\includegraphics[width=0.49\textwidth]{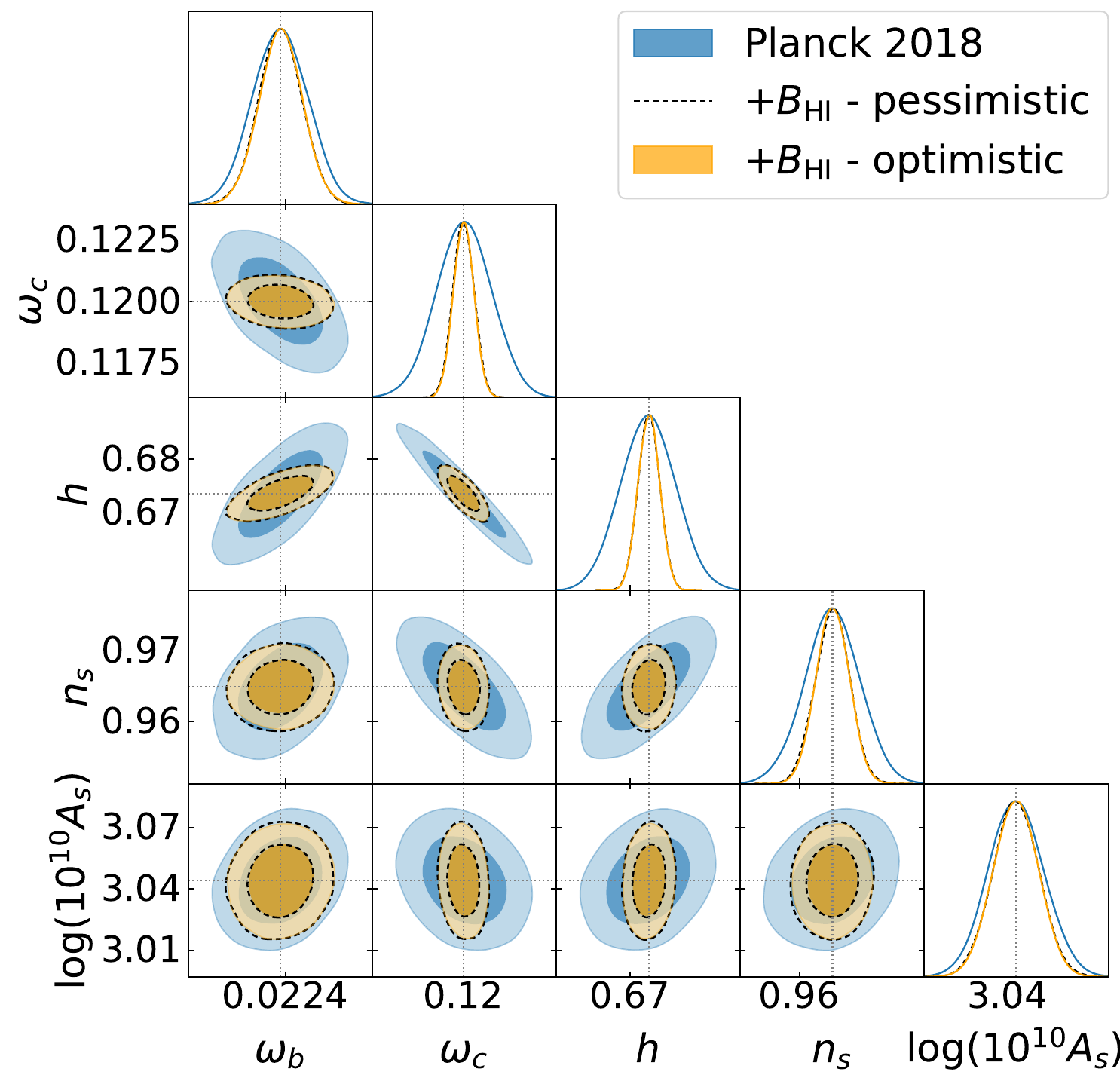}
\caption{ Same as Fig.~\ref{fig:HI_power_spectrum_forecasts} 
but for the HI bispectrum alone (left panel) or in combination with Planck data (right panel). Both panels show the pessimistic and optimistic foreground scenarios, corresponding to $k_{\parallel,\mathrm{min}} = 0.005,h,\mathrm{Mpc}^{-1}$ and $k_{\parallel,\mathrm{min}} = 0.001,h,\mathrm{Mpc}^{-1}$, respectively.}\label{fig:bispectrum_cosmo}
\end{figure}

Cosmological constraints are typically derived from the power spectrum. The inclusion of the \hi\ bispectrum, alongside the traditional two-point statistic, provides notable improvements in parameter forecasts by breaking degeneracies and enhancing precision, as demonstrated in recent analyses of galaxy survey data and simulations \citep{Philcox:2021kcw,Jung:2022gfa,Philcox:2022frc,Chudaykin_2025,Bakx:2025pop}.  Even under Gaussian initial conditions, higher-order correlators become non-zero due to the nonlinear evolution of structure under gravity. The leading higher-order statistic is the bispectrum- i.e. the Fourier transform of the three-point correlation function-which for \hi\ takes the form
\begin{equation}\label{eq:21_bispectrum}
\langle\delta T_{\hi}(\bk_1)\, \delta T_{\hi}(\bk_2)\, \delta T_{\hi}(\bk_3)\rangle = (2\pi)^3 \,\delta_{\rm D}(\bk_1 + \bk_2 + \bk_3)\, B_{\hi}(\bk_1, \bk_2, \bk_3),
\end{equation}
where the Dirac delta function enforces momentum conservation.

The \hi\ bispectrum is primarily employed for its significant constraining power on the primordial Universe and in the presence of non-Gaussian initial conditions, as described in \citet{Fonseca01.2026.SKA}. Furthermore, its late-time gravitational component serves as a powerful probe for extracting a wide range of cosmological information, as discussed in \citet{Majumdar01.2026.SKA}. In this section, we use the \hi\ bispectrum from SKA-Mid Band 1 to forecast constraints on the standard \LCDM\ cosmological parameters. Following \citet{Karagiannis:2022ylq}, we perform Fisher forecasts based on a tree-level model, which is restricted within the perturbative regime. The model incorporates redshift-space distortions (RSD) and a second-order bias expansion, while theoretical-error covariance terms account for the impact of neglected one-loop corrections. To assess the impact of foreground contamination, we also consider two large-scale cut in the bispectrum analysis. The full parameter set, encapsulating the main sources of uncertainty in our model, includes five cosmological parameters and eleven additional nuisance parameters per redshift bin\footnote{These are the HI IM bias coefficients, the AP multiplicative factors, the growth rate, the shot-noise terms and the FOG amplitudes, i.e. the full free parameters vector considered is $\bm{\theta}(z_i)=\Big\{\omega_{\rm b},\,\omega_{\rm c},\, h,\,\ln(10^{10}A_{\rm s}),\,n_{\rm s}\,;
D_A(z_i),H(z_i),f(z_i),b_1(z_i),b_2(z_i),b_{s^2}(z_i),\sigma_P(z_i),\sigma_B(z_i),P_\varepsilon(z_i),P_{\varepsilon\varepsilon_{\delta}}(z_i),B_{\varepsilon}(z_i)\Big\}.$, all treated as free.} After marginalising over the nuisance parameters, the resulting cosmological 1-$\sigma$ forecasts are presented in Fig. \ref{fig:bispectrum_cosmo}. Although the bispectrum results are not competitive with those from the power spectrum analysis (see Fig.~\ref{fig:HI_power_spectrum_forecasts}), they still provide a notable improvement when combined with Planck data. Moreover, the joint use of the power spectrum and bispectrum could further enhance the cosmological constraints achievable with SKAO \citep{Karagiannis:2022ylq,Randrianjanahary:2023rgp}. Finally, the impact of foreground cuts on the bispectrum-based forecasts is found to be minimal. We adopt a foreground-avoidance approach, excluding from the analysis all Fourier modes with $k_\parallel < k_\parallel^{\rm min}$, where radio foregrounds---dominated by spectrally smooth Galactic synchrotron and free-free emission---contaminate the long-wavelength radial modes~\citep{Liu2011,Shaw:2013wza,Shaw:2014khi}. We consider $k_\parallel^{\rm min} = 0.001\,h/{\rm Mpc}$ (optimistic) and $0.005\,h/{\rm Mpc}$ (pessimistic) to bracket the impact of foreground contamination on our forecasts.






\subsection{\hi\ stacking}
Emission line stacking is a powerful probe of the gas content of galaxies. Stacking measurements using \hi\ intensity maps have been made using the Parkes telescope \citep{2019MNRAS.489..385T,2020MNRAS.498.5916T}, the Canadian Hydrogen Intensity Mapping Experiment \citep{2023ApJ...947...16A}, and the MeerKAT telescope \citep{2025ApJS..279...19C}. The detection using the precursor MeerKAT telescope, as presented in \cite{2025ApJS..279...19C}, reveals that the \hi\ stacking can be used to constrain potential systematic effects as well as the \hi\ density of the Universe.

Following the data analysis pipeline described in \cite{2025ApJS..279...19C}, we generate mock observations of \hi\ signal. A catalogue of overlapping spectroscopic galaxies is also generated, and the \hi\ signal can then be stacked onto the positions of the galaxies. The signal is then processed into stacking measurement as described in \citet{Spinelli01.2026.SKA}. The \hi\ signal is then rescaled in each redshift so that the \hi\ density $\Omega_{\rm \hi}(z)$ follows the scaling relation described in \cite{2019MNRAS.489.1619H}. Furthermore, an oscillation systematic component, found in the data analysis of \cite{2025ApJS..279...19C}, is added to the mock simulation. The amplitude of the oscillating systematics is chosen to be $A_{\rm sys} = 0.1$ and the frequency $\nu_{\rm sys} = 20\,$MHz, which is motivated by observational data. A total of 100 realisations are generated to calculate the mean \hi\ signal as well as its covariance.

\begin{figure}
    \centering
    \includegraphics[width=1.0\textwidth]{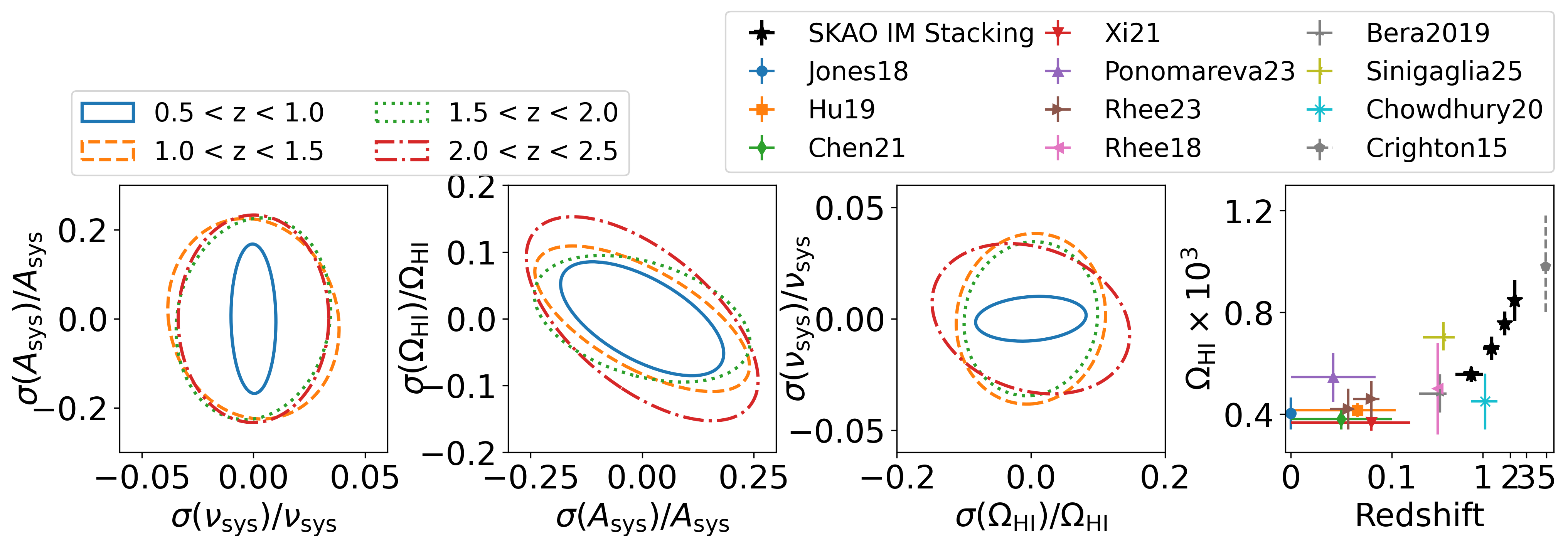}
    \caption{Forecasts for the constraints on the neutral hydrogen density and the potential observational systematics using emission line stacking of single-dish intensity mapping data with the \wideims. The first three panels from left to right show the forecasts on the fractional constraints on the model parameters. The rightmost panel shows the forecasts for the constraints on $\Omega_{\rm \hi}$ using stacking (``SKAO IM Stacking``) at $z\sim 0.5-2.5$. For comparison, a selection of current measurements using \hi\ emission line observations \citep{2018MNRAS.477....2J, 2018MNRAS.473.1879R, 2019ApJ...882L...7B, 2019MNRAS.489.1619H, 2020Natur.586..369C, 2021MNRAS.508.2758C, 2021MNRAS.501.4550X, 2023MNRAS.522.5308P, 2023MNRAS.518.4646R, 2025arXiv250611280S} are shown. SKAO is capable of probing the \hi\ density at $z>1$, which was thought to be only observable through the absorption line of damped Lyman alpha systems (e.g. \citealt{2015MNRAS.452..217C}).}
    \label{fig:stack}
\end{figure}

The simulation pipeline is then used to calculate the Fisher matrix by varying the model parameters, which then gives the forecasted constraints of the stacking measurement. In total, we model three parameters, the \hi\ density $\Omega_{\rm \hi}$, the amplitude of the oscillating systematics $A_{\rm sys}$ and the frequency of the systematics $\nu_{\rm sys}$. We choose a step size of 0.01 of the fiducial values to calculate the partial derivatives of the mean signal and the covariance, which are needed for the calculation of the Fisher matrix. The results are shown in \autoref{fig:stack}. Using the single-dish intensity mapping data from the Wide Band 1 survey, stacking measurements will be able to precisely constrain $\Omega_{\rm \hi}$ as well as the systematics. In particular, we find that $\sigma (\Omega_{\rm \hi}) / \Omega_{\rm \hi} = \{0.054, 0.069, 0.062, 0.095\}$ at $z\sim \{[0.5, 1.0],[1.0,1.5],[1.5,2.0],[2.0,2.5]\}$. The constraining power is comparable to the emission line observations using radio interferometry at $z\sim0.0$, with the unique ability of probing high redshifts up to $z\sim 2.5$.
\subsection{Dark Matter}

 With intensity mapping, `no photon is left behind'. That is, we collect all integrated emission from $\hi$; hence, we are sensitive to all sources containing cold gas, even the smallest dark matter haloes dense enough to host some, and that typically go undetected in standard observational techniques. This unique feature of $\hi$ intensity mapping makes it a powerful probe of cosmological scenarios in which the low-mass end of the halo mass function is modified compared to the standard $\Lambda$CDM model.

For instance, if dark matter is of particle origin, the perfectly `cold' assumption is an asymptotic one. Actually, the particle mass determines the streaming velocity and sets a specific cut-off in the halo mass function, that is, the minimum mass of existing dark matter haloes and their abundance \citep{2024MNRAS.52711740L}. In particular, and counter-intuitively, in universes in which structure formation is slightly suppressed, $\hi$ is forced to cluster in the most massive —and more biased— dark matter haloes, increasing the expected $\hi$ signal power spectrum \citep{Carucci2015}. 
Forecasts indicate that, assuming an intensity mapping survey with SKA1-LOW with an area of $\approx3-6$ deg$^2$ at $z = 3 - 5$, we will be able to rule out a 4 keV DM model with 5000 hours of observations, with a statistical significance larger than $3\sigma$ \citep{Carucci2015}.  In general, this is a unique observational feature of $\hi$ intensity mapping, which can be observed in any cosmological scenario that modifies structure formation \citep{Carucci2017}, turning a small-scale, hard-to-detect feature into a strong, large-scale effect on the $\hi$ intensity mapping power spectrum.

\section{Summary}
\hi\ Intensity Mapping provides a new independent tracer of the large scale structure measuring the cosmic \hi\ abundance and distribution through cosmic time $0 < z < 6$ with unprecedented precision.
The SKA1 Wide Band 1 Survey in combination with the Deep SKA1-LOW Survey will provide a new indispensable legacy dataset for \hi\ science and cosmology. 
If a commensal observing mode is employed, the SKA-Mid \wideims\ will be incredibly efficient in using the SKA-Mid telescope time and provide visibility data to produce radio continuum maps for a wealth of science cases as well as enabling slow transient science; more details are provided in \cite{Chatterjee01.2026.SKA}.
In this chapter we showcased the potential of the probe through a selection of forecasts on constraints on standard cosmology, using standard probes such as the \hi\ power spectrum, Baryon Acoustic Oscillations, the turnover scale, the bi-spectrum and \hi\ stacking. Furthermore, \hi\ IM will provide an excellent addition to cosmology with synergies, adding large scales and high redshift resolution over a continuous range of redshifts. More extensive forecasts including non-$\Lambda\rm CDM$ cosmologies and tests of General Relativity are presented in chapters \cite{Camera01.2026.SKA} and \cite{Fonseca01.2026.SKA}.
\bibliographystyle{abbrvnat-maxbibnames4}
\bibliography{chapter} 

\end{document}